\documentclass[12pt]{article}

\def\cN{{\cal N}}
\def\cD{{\cal D}}

\def\cV{{\cal V}}
\def\cA{{\cal A}}
\def\cB{{\cal B}}
\def\l{\lambda}
\def\z{\zeta}
\def\s{\sigma}
\def\ve{\varepsilon}

\def\ta{\tilde{a}}
\def\tx{\tilde{x}}

\def\g{\gamma}
\newcommand{\eqn}[1]{(\ref{#1})}
\def\beq{\begin{equation}}
\def\eeq{\end{equation}}

\begin {document}

\begin{titlepage}
\hfill HU-EP-00/57

\vspace{2cm}

\begin{center}
{\bf \Large
Brane--worlds in
5D supergravity}

\vspace{1cm}

Anna Ceresole $^\star$ and Gianguido Dall'Agata $^\sharp$

\vspace{1cm}

{$^\star$ Dipartimento di Fisica, Politecnico di Torino and \\
Istituto Nazionale di Fisica Nucleare, Sezione di Torino \\
C.so Duca degli Abruzzi, 24, I-10129 Torino.\\
{\tt ceresole@athena.polito.it}}

\medskip

{$^\sharp$ Institut f\"ur Physik, Humboldt Universit\"at \\
Invalidenstra\ss{}e 110, 10115 Berlin, Germany\\
{\tt dallagat@physik.hu-berlin.de}
}

\vspace{1cm}

\begin{abstract}
We summarise the present status  of supersymmetric
Randall--Sundrum brane--world scenarios and report on their
possible realisation within five--dimensional matter coupled
$\cN=2$ gauged supergravity. 
\end{abstract}

\end{center}

\vspace{3cm}

Proceedings of the talks given by G. Dall'Agata 
at the Supersymmetry and Quantum Field Theory
Conference, Kharkov, 25--29 July 2000
and at the European RTN Network conference, Berlin, 4--10 October 2000.

\end{titlepage}

The  idea of obtaining phenomenology from more than four
dimensional space--times is almost as old as Einstein's theory of
relativity. Physically sensible results can arise upon imposing
some confining mechanism for particles and interactions in the
extra dimensions. Many different such mechanisms have been
proposed over the years, and almost all of them have given new
insights on the possible structure and features required for any
consistent theory in $D> 4$.

The first proposal was the Kaluza--Klein recipe \cite{KK}, where
all the extra dimensions are compactified on some internal
manifold, such that vibrations over this space are seen as masses
and charges for particles from the four--dimensional point of
view. This old idea, first suggested to unify electrodynamics and
gravity, is still much used nowadays in strings and
supergravities  to extract phenomenology from these ten and
eleven--dimensional theories by compactifying them on internal
manifolds of the size of the Planck length \cite{sezgin}.

A new step forward was provided by the Ho$\check{r}$ava--Witten
model \cite{HW} (and its five-dimensional realisation \cite{LOSW}).
In this model, particles and fields of spin
ranging from 0 to 1 are confined to membranes embedded in an
eleven--dimensional space with still a compact eleventh direction.
This new mechanism  opens the possibility to obtain at least one
extra dimension of size greater than the usual Planck scale.

Although the constraining on branes is a valid mechanism for spin
0 and 1 fields, one cannot use it for gravity. It is indeed known
that gravity couples to all energy sources,  and aiming
essentially at describing the geometry of the whole space, it cannot be
confined to a submanifold only.

Actually, the idea of viewing our universe as some membrane
embedded in a higher--dimensional space can be traced back to
\cite{Regge}, but for what just said it was left undeveloped  at
that time.

So, how is it possible to confine spin $3/2$ and spin 2 fields on membranes?
In spite of the above objections, last year Randall and Sundrum
\cite{RS1} have proposed a mechanism which goes very close to
constraining gravity. They show that one can obtain some graviton
bound state which is centered and concentrated on a
four--dimensional subspace of a five--dimensional bulk.
The main new feature of this scenario is that they actually
impose the ambient space to have a non--vanishing cosmological
constant such that a new mechanism for confining gravity
fluctuations can occur. An outstanding result is that the extra
dimension involved  can be made arbitrarily large and even
completely unfolded, without ruining the possibility of having
confinement. This implies that one could obtain effective
four--dimensional gravity  starting from a  non--compact
five--dimensional space--time.

Since this scenario is claimed to solve many longstanding
phenomenological problems, like the cosmological constant value
and the hierarchy problem, it seems compelling to find out
whether it can be embedded into higher dimensional supersymmetric
theories like supergravity and string theory, that provide the
most natural candidates for  field unification. This
would provide for them a more rigorous derivation while avoiding the
need of many fine--tunings and the ad-hoc selection of their
characteristics.

First of all we would like to distinguish between three different
setups which are all called Randall--Sundrum models, but actually
have different theoretical relevance.

In the original proposal \cite{RS1}, meant to provide a
solution to the hierarchy problem, there are two membranes
located at the orbifold fixed points of a still compact fifth
dimension. The complete action includes 5D gravity and sources
for the two membranes:
$$
S = S_{bulk}+S_{branes}.
$$
One serious drawback of the {\sl two--brane scenario} was the presence of a
negative tension membrane that however, as was later shown, could
be consistently taken to infinity. Thus came the  second setup,
the {\sl one-brane scenario} \cite{RS2}  with only one  singular `thin'
(delta-functional) membrane, where gravity is confined by a
volcano potential, surrounded by two slices of AdS space.
The metric of such configuration can be put in the form
\beq
ds^2=a^2(y) \, dx^2 + dy^2,
\label{met}
\eeq  respecting four--dimensional
Poincar\'e invariance, and where the {\sl warp factor} behaves for $|y| \to \infty$ as
$a^2(y)=e^{2A(y)}\sim e^{-k|y|}$ ($k>0$).

Due to the infinite extension of the fifth dimension $y$, this
model leads to the very intriguing possibility of getting a real
substitute for the usual compactification scheme. Therefore this
{\it alternative to compactification} is  not only appealing
from a phenomenological point of view, but also from a
purely theoretical one, since one could try to implement it in
supergravity and string theory, replacing the usual Kaluza--Klein
mechanism.

Making any of these scenarios compatible with some supersymmetric theory
has been attempted in various ways \cite{susy1,StelleLiu} and ultimately
requires the implementation of supersymmetry on singular spaces
\cite{BKVP}. This state of affairs is still somewhat unsatisfactory,
even in the example where you can eliminate the negative tension brane.

The obvious improvement would be to trade  the singular
source with a smooth (``thick'') domain wall solution of some
(super--)gravity theory \cite{thick}, the {\sl thick brane scenario}.
 In this way our universe
could be seen as some natural result of the dynamics of one of
the known theories in higher--dimensional space--time. More
precisely, due to the presence of a cosmological constant, the
theories to be explored for the embedding of such scenarios are
the gauged supergravities, and in the simplest instance the
five--dimensional ones \cite{GuSiTo,AnnaGianguido}. Unhappily, all
the many attempts made up to now in this direction
can  be summarised
in a no--go theorem \cite{KL} stating that no smooth
Randall-Sundrum scenario can be found within 5D gauged
supergravity interacting with an arbitrary number of vector and/or
tensor multiplets. Moreover, even when also hypermultiplets are
coupled, one encounters various difficulties that, although not
in the form of a no--go theorem, hint to the fact that if trapping
of gravity on the brane in 5D supergravity is possible, it
certainly must take place in a very specific model. The reason
for such rareness stems on the requirement of having a
supersymmetric flow joining two stable IR fixed points with the
same cosmological constant, lying on each side of the brane. Upon
examining a large class of supergravity scalar potentials, these
constraints seem to be quite hard to meet.

Possible ways out of this picture are the inclusion of massive
multiplets, as suggested by models containing the ``breathing mode''
\cite{mirjam} or other generalisations of the existing
supersymmetric theories, that are presently under investigation.

\section{$\cN = 2$, $D = 5$ gauged supergravity}

It is well known that {\sl gauged} supergravities, where some of
the global isometries of the standard theory ($R$--symmetry
included) are made local, admit a scalar field potential that, at
some critical point, generates the cosmological term.
More precisely, the global isometries of the scalar manifold are
made local by substituting the ordinary derivatives  in the
Lagrangean with new covariant derivatives, depending on the vector
fields which become gauge bosons of this invariance
$\partial_\mu \longrightarrow {\cal D}_\mu = \partial_\mu + g A_\mu$.
As this process obviously brakes supersymmetry, one has to
restore it by adding new terms (shifts), of first order in the
coupling constant $g$, in the supersymmetry transformation rules.
This will  create new contributions in the Lagrangean that are also of
first order in $g$, and are interpreted as mass terms for the
Fermi fields. To cancel supersymmetry variations of these terms,
one has to add to the Lagrangean a further piece of order $g^2$:
the scalar potential. The great miracle of  gauged
supergravities is that, under certain conditions, the process does not
go on to infinity as
all the higher order terms vanish identically, leaving with a consistent
supersymmetric theory\footnote{In a geometric
approach, the change in the derivative reflects
into a direct modification of the Bianchi identities for the
various fields, and the modification of the
supersymmetry rules is a natural consequence of their closure.}.

Given that five dimensions are necessary for an effective
four--dimensional brane--world, it is natural to look  first into
the minimal supersymmetric $\cN = 2$ ({\sl i.e.} with eight real
supercharges) supergravity, and in order to avoid the
no-go theorem \cite{KL}, one considers the generic interaction with
an arbitrary number $n_V$ of vector, $n_T$ of tensor and $n_H$ of
hypermultiplets \cite{AnnaGianguido}. The fermionic fields of
this model  are two gravitini $\psi_\mu^i$ that are symplectic
Majorana spinors ($i=1,2$ are $SU(2)_R$ indices
and $\mu=0,\ldots,4$), $n_V+n_T$ gaugini
$\lambda^{i{\ta}}$ (${\ta}=1,\ldots,n_V+n_T$), and the hyperini
$\zeta^A$ with $A=1,\ldots,2n_H$. The bosonic sector is composed,
beside the graviton $e^a_\mu$, the graviphoton $A_\mu^0$, the
$n_V$ vectors $A_\mu^I$ and $n_T$ tensors $B_{\mu\nu}^M$, by the
two sets of scalars $\phi^{\tilde x}$ and $q^X$ belonging to the
vector and hyper--multiplets, spanning respectively a very special
 and a quaternionic target space.

Quoting only what is essential to search for smooth RS solutions,
the scalar potential is given by
\beq
\label{scalpot}
\cV(\phi,q)=g^2 \{
2W^{\ta}W^{\ta} - \left[ 2 P_{ij} P^{ij} -P_{\ta ij}P_{\ta}^{ij}
 \right]+2\cN_{iA}\cN^{iA}\}
\eeq
where the various quantities are the  shifts
appearing in the supersymmetry rules for the fermion fields,
whose relevant parts (in a backgound with vanishing fermions and
vectors) are
\begin{eqnarray}
\label{susyrule1}
\delta_{\ve} \psi_{i\mu} &=& \cD_\mu\ve_i
+\frac{i}{\sqrt{6}} g \,  \g_\mu \ve^j P_{ij}, \\
\label{susyrule2}
\delta_{\ve} \l_i^{\ta} &=& - \frac{i}{2} f^{\ta}_{\tx} \g^\mu
\ve_i \, {\cD}_\mu \phi^{\tx} +
g \ve^j P^{\ta}_{ij} + g  W^{\ta} \ve_i,  \\
\label{susyrule3}
\delta_\ve \z^A &=& - \frac{i}{2}f_{iX}^A \g^\mu \ve^i
{\cD}_\mu q^X +  g  \ve^i \cN_{i}^A.
\end{eqnarray}

The above symbols are defined as
\begin{eqnarray}
\label{pij} &P_{ij}\equiv h^I P_{I\, ij},\qquad\qquad
P^{\ta}_{ij} \equiv h^{\ta I} P_{I\, ij}, \\
&W^{\ta}  = \frac{\sqrt{6}}{4} h^I K_I^{\tx} f_{\tx}^{\ta},\qquad\qquad
 \cN^{iA}= \frac{\sqrt{6}}{4} h^{I} K^X_{I} f_{X}^{Ai}.
\end{eqnarray}
in terms of various geometric scalar dependent quantities such as
$h^I(\phi),h^{\ta I}(\phi)$, the vielbeins of the very special
and quaternionic manifolds $f^{\ta}_{\tx}(\phi)$, $f^A_{iX}(q)$, the
corresponding  Killing vectors $K^{\tx}_I$ and $K^X_I$, and the
quaternionic prepotential $P_{I\, ij}(q)$ \cite{GuSiTo,AnnaGianguido}.
Notice that the $W^{\ta}$ shift in the gluino susy rule is specifically
due to the tensor multiplet couplings \cite{GuSiTo}.

In order to search for supersymmetric configurations, one solves
Killing spinor equations that are obtained by setting to zero
supersymmetry variations of the Fermi fields. Moreover, these
latter also give the supersymmetry flow equations, that are first
order differential equations giving rise to
solutions of the full equations of motion.

\section{RS solutions}

Solitonic solutions of the theory at
hand, with suitable features to yield an RS scenario, must have
a metric of the form \eqn{met}.
Moreover, the scalar potential \eqn{scalpot} must have at
least two different critical points, where the scalar fields $\Phi=\{\phi,q\}$
reach some fixed value $\Phi \to \Phi^*$, such that as $|y| \to \infty$, the
metric approaches that of an AdS space,
$ds^2 \to e^{-|y|} dx^2 + dy^2$.
Therefore the warp factor must approach zero as we go toward this fixed
point.

To summarise the outcome of the analysis carried out when only vector multiplets are
coupled to supergravity \cite{KL},
the supersymmetry flow equations around the fixed points yield the `wrong'
behaviour for the fields as functions of the warp factor,
\beq
\phi^{\tx}(a) = \phi^{{\tx}*} + \frac{c^{\tx}}{a^2}.
\eeq
Thus, the warp factor turns out to be {\it always increasing} as we
move towards the fixed point, while it should decrease for  an RS
solution.

The obvious hope is that things could change if we allow for couplings
to hypermultiplets.

Let us then examine the susy equations \eqn{susyrule1}--\eqn{susyrule3}.
For BPS solutions to exist, one has to
find some Killing spinors which, in addition to the usual
constraint coming from setting to zero the gravitino susy rule \eqn{susyrule1}
\beq
\label{cond1}
i \g^y \ve_{i} \sim g P_{ij} \ve^j,
\eeq
also satisfy
\beq
\label{cond2}
\delta_\ve \l_i^{\ta}={{\cA^{\ta}}_i}^j \ve_j = 0
\eeq
and
\beq
\label{cond3}
f^X_{iA}\delta_\ve \z^A={{\cB^{X}}_i}^j \ve_j = 0
\eeq
where the operator matrices $\cA^{\ta}$ and $\cB^X$ can be read off from
\eqn{susyrule2} and \eqn{susyrule3} respectively.

The key point is that, once  the solution to
\eqn{cond1} is inserted into \eqn{cond2} and \eqn{cond3}, the $\cA$ and $\cB$
operators reduce to simple SU(2) matrices acting only on the SU(2) index of the Killing
spinor. This implies that the possibility of finding BPS solutions amounts to
having a Killing spinor eigenvector of  $\cA^{\ta}$ (the same is true for
$\cB^X$) with zero eigenvalue, i.e. the matrix $\cA^{\ta}$ must be degenerate.

A closer look shows that, after making explicit the $SU(2)$ structure, these matrices
actually have the same form:
\begin{eqnarray}
{{\cA^{\ta}}_i}^j &\sim& i S^{r\ta} (\s_r)_i{}^j + Q^{\ta}
\delta_i^j, \\
{{\cB^{X}}_i}^j &\sim& i S^{r X} (\s_r)_i{}^j + Q^{X}
\delta_i^j,
\end{eqnarray}
where  $S^{r\ta},S^{rX},Q^{\ta},Q^X$ indicate some real combination of the various
$y$--dependent quantities.
Requiring  det$\cA^{\ta} = 0$ (the same for $\cB$), imposes for each ${\ta}$
\beq
(Q^{\ta})^2 + (S^{\ta})^2 = 0,
\eeq
which has no solutions except for
\beq
Q^{\ta} = 0 = S^{\ta}.
\eeq

These conditions confirm that no RS scenario can be obtained
if tensor multiplets are present, since $Q^{\ta}\equiv W^{\ta}$ is precisely
the contribution of tensor multiplets to the gaugino susy rule
and thus to the potential.
All the other constraints instead lead to some partial
differential equations with respect to the $y$ coordinate involving the scalar
fields and the warp factor.
Such equations may or may not yield appropriate solutions, but cannot be
solved without selecting a specific model and computing the relevant
geometric quantities. Therefore, the study of the general supersymmetric
flow equations does not suffice to clear up the situation,
 as was the case for vectors and/or tensor multiplets
coupled to supergravity.

More recent studies \cite{inprep} seem to show that there is at least 
the possibility
of finding infra--red fixed points from the hypermultiplet scalar manifold.
Indeed, one can find critical points near which the scalars  behave as
\beq
\Phi^i \sim \Phi^{i*} + c^i a(y)^n,
\eeq
for some power $n>0$.
However, this is not enough to give an RS solutions, since we have seen
that one needs {\sl two} such points in the same theory, where the
value of the potential is the same, and so far none of these have been found.

\section{Outlook}

We have discussed the simplest and most natural supersymmetric framework where smooth
Randall--Sundrum  brane worlds could fit, that is the fully
coupled $D = 5$, $\cN = 2$ gauged supergravity. It is perhaps  interesting
to notice that although unsuccessful up to now, the search for such cosmological
scenarios has already stimulated new results. Indeed, it was among the main
motivations for constructing the above theory, that was previously known
only in absence of hypermultiplets \cite{GuSiTo}.

Further searches bring to the more thorough exploration of 
specific hyper--matter coupled models, or even to more general extensions.
For instance, beside considering more systematically theories 
with a higher number
of supersymmetries, that would change the susy rules, one should look at new
possible couplings of the $\cN = 2$ theory to massive multiplets that exist as
representations of the $SU(2,2|1)$ supergroup.

A promising  possibility seem to be to add  the interaction with massive
{\sl vector} multiplets,
since these should be linked to scalars with massess big enough to
allow for the appearence of infra--red critical points 
\cite{StelleLiu, inprep}.
The final word is likely to be hidden in the general discussion of $d
= 5$ supersymmetric flows/domain--walls, that is now under investigation 
\cite{inprep}.

\vskip0.5cm
\noindent
{\large \bf Acknowledgements}

\smallskip
\noindent This work is supported in part by European Commission
under TMR project HPRN-CT-2000-00131, with A. C. associated to
Torino University.


\end{document}